\documentclass[12pt]{iopart}
\usepackage{color}

\usepackage{graphicx}
\begin{document}
\letter{Superconductivity at 35 K by self doping in RbGd$_2$Fe$_4$As$_4$O$_2$}

\author{Zhi-Cheng Wang$^1$, Chao-Yang He$^1$, Si-Qi Wu$^1$, Zhang-Tu Tang$^1$, Yi Liu$^1$, Abduweli Ablimit$^1$, Qian Tao$^1$, Chun-Mu Feng$^1$, Zhu-An Xu$^{1, 2, 3}$, and Guang-Han Cao$^{1, 2, 3}$}
\address{$^1$ Department of Physics, Zhejiang University, Hangzhou 310027, People's Republic of China}
\address{$^2$ State Key Lab of Silicon Materials, Zhejiang University, Hangzhou 310027, People's Republic of China}
\address{$^3$ Collaborative Innovation Centre of Advanced Microstructures, Nanjing 210093, People's Republic of China}

\ead{ghcao@zju.edu.cn}

\begin{abstract}

We report synthesis, crystal structure and physical properties of a novel quinary compound RbGd$_2$Fe$_4$As$_4$O$_2$. The new iron oxyarsenide is isostructural to the fluo-arsenide KCa$_2$Fe$_4$As$_4$F$_2$, both of which contain separate double Fe$_2$As$_2$ layers that are self hole-doped in the stoichiometric composition. Bulk superconductivity at $T_\mathrm{c}$ = 35 K is demonstrated by the measurements of electrical resistivity, dc magnetic susceptibility and heat capacity. An exceptionally high value of the initial slope of the upper critical field ($\mu_0$d$H_{\mathrm{c2}}$/d$T|_{T_\mathrm{c}}$ = 16.5 T/K) is measured for the polycrystalline sample.


(Some figures may appear in colour only in the online journal)

\end{abstract}



\submitto{\JPCM}
\maketitle
Since the discovery of high-$T_\mathrm{c}$ superconductivity in iron-based compounds\cite{hosono,cxh}, great efforts have been devoted to expanding the family of Fe-based superconductors (FeSCs)\cite{hosono-pc,cxh-scm}. Structurally, all the FeSCs discovered to date contain two-dimensional Fe$_2$$X_2$ ($X$ = As or Se) layers, crystallizing in different crystal-structure types\cite{jh}. The relatively simple structures include (i) 11-type FeSe\cite{wmk}, (ii) 111-type LiFeAs\cite{jcq}, (iii) 122-type (Ba$_{1-x}$K$_x$)Fe$_2$As$_2$\cite{johrendt}, (iv) 122*-type K$_x$Fe$_{2}$Se$_2$\cite{cxl}, and (v) 1111-type LaFeAsO$_{1-x}$F$_x$\cite{hosono}. There are also complex structures with perovskite-like blocks serving as the spacer layers\cite{32522-whh,42622P,ogino1,ogino2}. In recent years, novel-structure-bearing FeSCs were discovered continually\cite{nohara,cava,syl,42214,112-jpsj,112-jacs,cxh-nm}, which have brought growing vitality in the related research\cite{johnston,cxh-nsr}.

We proposed a rational route to new FeSCs via a strategy of structural design\cite{jh}. Nine structures were suggested to host superconductivity potentially. One of the candidate structures, exemplified as 1144-type ``KLaFe$_4$As$_4$", was recently realized in $AkAe$Fe$_4$As$_4$ ($Ak$ = K, Rb, Cs; $Ae$ = Ca, Sr) which show superconductivity at 31-36 K\cite{1144}. Additional 1144-type members containing rare-earth element Eu, namely $Ak$EuFe$_4$As$_4$ with $Ak$ = Rb and Cs, were subsequently synthesized\cite{Eu1144,RbEu1144,CsEu1144}, in which not only superconductivity but also full ferromagnetism (in the Eu sublattice) emerge\cite{RbEu1144,CsEu1144}. Following our structure-design strategy, very recently, we also succeeded in realizing another candidate structure in a series of quinary iron fluo-arsenides \textit{Ak}Ca$_2$Fe$_4$As$_4$F$_2$ (\textit{Ak} = K, Rb, Cs)\cite{wzc,wzc-scm}. These new 12442-type FeSCs, superconducting at about 30 K, are the first class of iron pnictides that contains separate double Fe$_2$As$_2$ layers.

As a matter of fact, the above two new structures belong to the category of structural intergrowth, similar to the multi-element FeSC Ba$_2$Ti$_2$Fe$_2$As$_4$O discovered earlier\cite{syl}. $AkAe$Fe$_4$As$_4$ can be viewed as an intergrowth of $Ak$Fe$_2$As$_2$ and $Ae$Fe$_2$As$_2$, while \textit{Ak}Ca$_2$Fe$_4$As$_4$F$_2$ (\textit{Ak} = K, Rb, Cs) is an intergrowth of 1111-type CaFeAsF and 122-type \textit{Ak}Fe$_2$As$_2$ (\textit{Ak} = K, Rb, Cs). As expected, lattice match is very crucial for obtaining an intergrowth compound\cite{jh,1144}. To expand the 12442-type superconducting family, therefore, one should first consider a good lattice match between the constituent crystallographic block layers. Another consideration is to find a higher $T_\mathrm{c}$ as far as possible. Note that there are 1111-type iron oxyarsenides $Re$FeAsO ($Re$ = rare-earth elements) whose lattice parameters $a$ are close to those of $Ak$Fe$_2$As$_2$, and additionally, doped GdFeAsO shows a record high $T_\mathrm{c}$ of 56 K in bulk FeSCs\cite{wc2008}, we thus chose $R$ = Gd and $Ak$ = Rb (the $a$ axes of GdFeAsO and RbFe$_2$As$_2$ are 3.915 {\AA}\cite{wc2008} and 3.863 {\AA}\cite{A122}, respectively) to explore a 12442-type \emph{oxyarsenide} FeSCs. In this Letter, we report synthesis, crystal structure and physical properties of the target compound RbGd$_2$Fe$_4$As$_4$O$_2$. The new material is hole doped without extrinsic chemical doping, which exhibits bulk superconductivity at $T_\mathrm{c}$ = 35 K, the highest $T_\mathrm{c}$ among hole-doped \emph{oxyarsenide} FeSCs.

The target material RbGd$_2$Fe$_4$As$_4$O$_2$ was synthesized by a solid-state reaction method, similar to our previous report\cite{wzc}. The source materials were Rb ingot (99.75\%), Gd ingot (99.9\%), Gd$_2$O$_3$ powder (99.9\%), Fe powders (99.998\%) and As pieces (99.999\%). Gd$_2$O$_3$ was heated to 1173 K for 24 hours in air so as to remove adsorbed water. The intermediate products of GdAs, FeAs and Fe$_2$As were prepared by direct solid-state reactions of their constituent elements at 1023 K for 15 hours. We also prepared an additional intermediate material ``Rb$_{1.03}$Fe$_2$As$_2$" by reacting Rb metal and FeAs at 923 K for 10 hours. After that, mixtures of RbFe$_2$As$_2$, GdAs, Gd$_2$O$_3$, FeAs and Fe$_2$As in the stoichiometric ratio were pressed into pellets, and then loaded in an alumina tube. To avoid the chemical reactions with quartz tubes, a Ta tube was jacketed before putting the sample into an quartz ampoule. This sample-loaded evacuated quartz ampoule was sintered at 1238 K for 40 hours, and then cooled down by switching off the furnace. The product was found to be stable in air.

Powder x-ray diffraction (XRD) was carried out on a PANAlytical x-ray diffactometer (Empyrean) with a CuK$_{\alpha1}$ monochromator (Johansson 1$\times$Ge111 Cu/Co) at room temperature. The collected XRD data ($20^{\circ}\leq 2\theta \leq 150^{\circ}$) were used for a Rietveld refinement. The structural refinement was based on the 12442-type structural model\cite{wzc}, using the software Rietan-FP\cite{rietan}. The occupation factor of each atom was fixed to 1.0. The refinement easily converges, which yields a weighted reliable factor of $R_\mathrm{wp}$ = 3.05\% and a goodness-of-fit of $S$ = 1.12, indicating the high quality of the refined crystallographic data.

The physical property measurements were done on a physical property measurement system (Quantum Design, PPMS-9) and a magnetic property measurement system (Quantum Design, MPMS-XL5). We employed a standard four-electrode method and the ac transport option for the electrical resistivity measurement. For the Hall measurement, four electrodes were made on a thin-square sample, forming a cross-shaped configuration. The excitation current was 20 mA, and the magnetic field swept from $-$80 to 80 kOe. The heat capacity was measured by a thermal relaxation method using a square-shaped sample plate (22.5 mg). The dc magnetic susceptibility was measured using a sample rod whose demagnetization factor is estimated to be 0.12(1).

\begin{figure}
\center
\includegraphics[width=12cm]{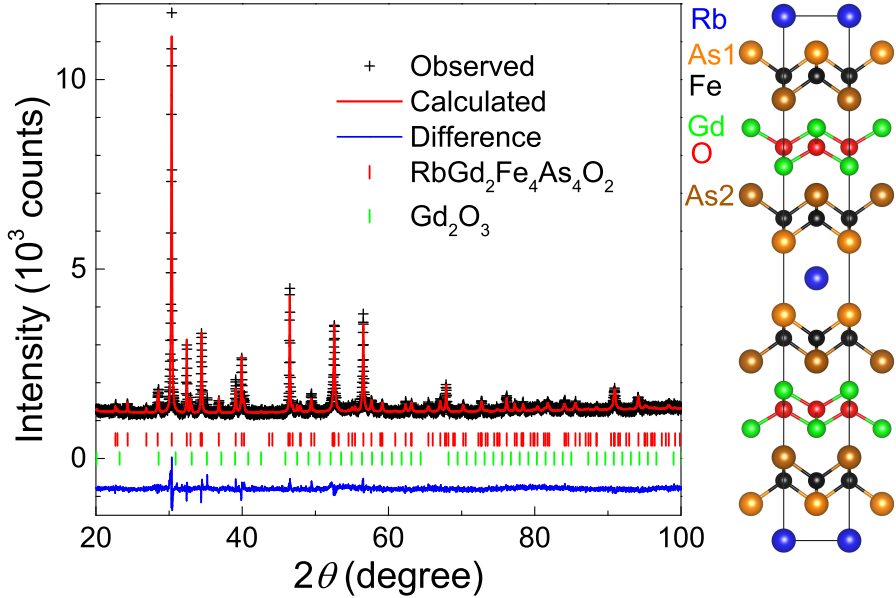}
\caption{\label{XRD} X-ray diffraction pattern and its Rietveld refinement profile (data not shown for $100^{\circ}\leq 2\theta \leq 150^{\circ}$ where the reflections are weak) for RbGd$_2$Fe$_4$As$_4$O$_2$. Shown on the right side is the crystal structure projected along the [100] direction.}
\end{figure}

Figure~\ref{XRD} shows the XRD pattern of the as-prepared RbGd$_2$Fe$_4$As$_4$O$_2$ sample. Most of the reflections can be well indexed with a 12442-type lattice ($I4/mmm$, $a \approx$ 3.89 {\AA}, $c \approx$ 31.3 {\AA}). The remaining unindexed weak reflections come from unreacted Gd$_2$O$_3$. We thus make a two-phase Rietveld analysis. The result shows that the mass fraction of the main phase RbGd$_2$Fe$_4$As$_4$O$_2$ is 94\% after a microabsorption correction\cite{rietan}.

Let us first examine changes of the crystal structure owing to the intergrowth between RbFe$_2$As$_2$ and GdFeAsO. The $a$ axis (see Table~\ref{structure}) is nearly the same as the average (3.889 {\AA}) of those of GdFeAsO\cite{wc2008} and RbFe$_2$As$_2$\cite{A122}. Meanwhile, the $c$ axis is close to the expected value ($2c_{\mathrm{GdFeAsO}}+c_{\mathrm{RbFe2As2}} =$ 31.317 {\AA}) of the structural intergrowth. Nonetheless, there are atomic adjustments in each building blocks. While the ``GdFeAsO" block become slender ($a$: 3.915 {\AA} $\rightarrow$ 3.901 {\AA}; $c$: 8.435 {\AA} $\rightarrow$ 8.538 {\AA}), the ``RbFe$_2$As$_2$" block goes the opposite ($a$: 3.863 {\AA} $\rightarrow$ 3.901 {\AA}; $c$: 7.224 {\AA} $\rightarrow$ 7.134 {\AA}). The structural reconstructions reflects charge redistribution between the two building blocks. As a result, the apparent Fe valence in RbGd$_2$Fe$_4$As$_4$O$_2$ becomes 2.25+, which is a mean value of 2+ in GdFeAsO and 2.5+ in RbFe$_2$As$_2$, leading to a self doping level of 0.25 holes/Fe.

Table~\ref{structure} compares crystal structures of the two 12442-type materials. The lattice parameters $a$ and $c$ of the oxyarsenide are 0.84\% and 1.1\% larger, respectively, due to the incorporations of larger ions of Rb$^{+}$ and O$^{2-}$. KCa$_2$Fe$_4$As$_4$F$_2$ is structurally featured by the obviously unequal Fe$-$As bond diantances\cite{wzc}. Similar case is seen in RbGd$_2$Fe$_4$As$_4$O$_2$. Nonetheless, the As height from the Fe plane and the band angles of As1$-$Fe$-$As1 and As2$-$Fe$-$As2, which are considered to be the relevant structural parameters that control $T_\mathrm{c}$\cite{mizuguchi,zhaoj,lee}, show that the values of RbGd$_2$Fe$_4$As$_4$O$_2$ is closer to the ideal one that generates the highest $T_\mathrm{c}$. This could serve an explanation for a higher $T_\mathrm{c}$ in RbGd$_2$Fe$_4$As$_4$O$_2$ (see below).

\begin{table}
\caption{\label{structure}
Crystallographic data of RbGd$_2$Fe$_4$As$_4$O$_2$ in comparison with those of  KCa$_2$Fe$_4$As$_4$F$_2$\cite{wzc}. The space group is $I4/mmm$ (No. 139). The atomic coordinates are as follows: K/Rb 2$a$(0, 0, 0); Ca/Gd 4$e$(0.5, 0.5, $z$); Fe 8$g$ (0.5, 0, $z$); As1 4$e$(0.5, 0.5, $z$); As2 4$e$(0, 0, $z$);  F/O 4$d$(0.5, 0, 0.25).}
\center
\begin{tabular}{lcccc}
\hline
  & RbGd$_2$Fe$_4$As$_4$O$_2$ &  &KCa$_2$Fe$_4$As$_4$F$_2$& \\
\hline
Lattice Parameters&  & & & \\\hline
$a$ (\r{A})  & 3.9014(2)&  &3.8684(2)& \\
$c$ (\r{A})  & 31.343(2)&  &31.007(1)& \\
$V$ (\r{A}$^{3}$)   & 477.06(4)&  &463.99(3)&  \\
\hline
Coordinates ($z$)& &  &  & \\ \hline
Ca/Gd &0.2138(1)&  &0.2085(2)  & \\
Fe &0.1138(2)&  &0.1108(1)  &\\
As1 &0.0697(2)&   &0.0655(1) &\\
As2  &0.1591(2)&   &0.1571(1) & \\
\hline
Bond distances&& &&  \\\hline
Fe$-$As1 (\r{A}) & 2.391(6) &   & 2.390(3) &  \\
Fe$-$As2 (\r{A}) &  2.413(6) & & 2.409(3) & \\
\hline
As Height from Fe Plane&& &&  \\\hline
As1 (\r{A}) & 1.382(13) &   & 1.405(3) &  \\
As2 (\r{A}) &  1.420(13) & & 1.436(3) & \\
\hline
Bond Angles && &&  \\\hline
As1$-$Fe$-$As1 ($^{\circ}$) & 109.4(3) &   & 108.0(2) &  \\
As2$-$Fe$-$As2 ($^{\circ}$) & 107.9(3) & & 106.8(2) &   \\
\hline
\end{tabular}
\end{table}

\begin{figure}
\center
\includegraphics[width=8cm]{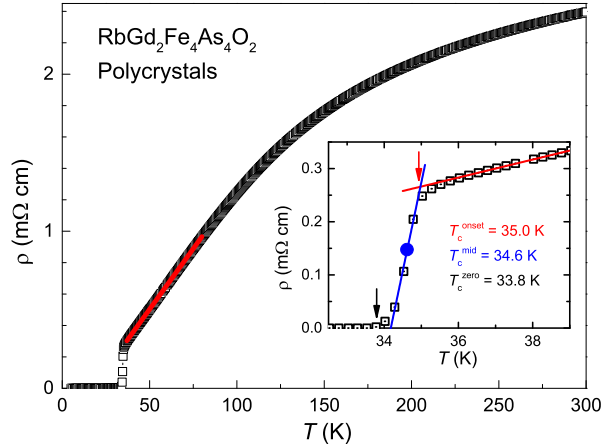}
\caption{\label{rt} Temperature dependence of resistivity for the RbGd$_2$Fe$_4$As$_4$O$_2$ polycrystalline sample. The inset magnifies the superconducting transition, in which the onset, midpoint, and zero-resistance temperatures are defined.}
\end{figure}

Figure~\ref{rt} shows temperature dependence of resistivity [$\rho(T)$] of RbGd$_2$Fe$_4$As$_4$O$_2$. Although the $\rho(T)$ data were obtained from polycrystalline samples, empirically they may represent the in-plane resistivity behavior for FeSCs. The $\rho(T)$ curve indicates a metallic conduction with a broad humpback at $\sim$ 150 K. Such a humpback is frequently observed in hole-doped FeSCs\cite{whh,johrendt,wzc}. Similar phenomenon in heavily hole-doped \textit{Ak}Fe$_2$As$_2$ (\textit{Ak} = K, Rb and Cs) is explained as an incoherent-to-coherent crossover in relation with an emergent Kondo lattice effect\cite{wutao}. The $\rho(T)$ data show a linear behaviour from $\sim$75 K to 35 K below which a sharp superconducting transition shows up. The transition width (defined by the temperature interval of the resistivity drop from 90\% to 10\%) is only 0.6 K, albeit of polycrystalline samples. The midpoint and zero-resistance temperatures ($T_{\mathrm{c}}^{\mathrm{mid}}$ and $T_{\mathrm{c}}^{\mathrm{zero}}$) are 34.6 and 33.8 K, respectively, as shown in the inset of  Fig.~\ref{rt}.

\begin{figure}
\center
\includegraphics[width=8cm]{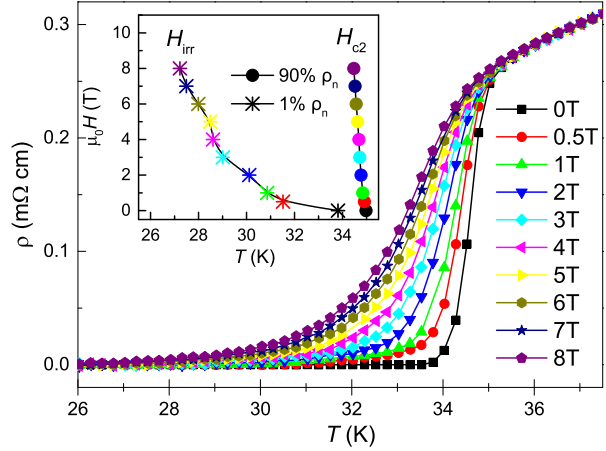}
\caption{\label{hc2} Superconducting transitions under external magnetic fields for RbGd$_2$Fe$_4$As$_4$O$_2$. The inset shows the superconducting phase diagram derived.}
\end{figure}

The superconducting resistive transitions under external magnetic fields are shown in Fig.~\ref{hc2}. With increasing magnetic field, $T_{\mathrm{c}}^{\mathrm{onset}}$ decreases slowly, but $T_{\mathrm{c}}^{\mathrm{zero}}$ decreases rapidly. We employed conventional criteria to extract the upper critical fields [$H_{\mathrm{c2}}(T)$] and the irreversible field [$H_{\mathrm{irr}}(T)$]. Namely, $T(H_{\mathrm{c2}})$ and $T(H_{\mathrm{irr}})$ are defined as the temperatures at which the resistivity drops to 90\% and 1\% of the extrapolated normal-state value. The derived superconducting phase diagram is depicted in the inset of Fig.~\ref{hc2}. $H_{\mathrm{c2}}(T)$ is almost linear with a slope of $|\mu_0$d$H_{\mathrm{c2}}$/d$T|$ = 16.5$\pm1$ T/K, which is about three times large as those of 1144-type CaKFe$_4$As$_4$\cite{canfield1}, RbEuFe$_4$As$_4$\cite{RbEu1144} and RbEuFe$_4$As$_4$\cite{CsEu1144}. In general, the $H_{\mathrm{c2}}(T)$ data for polycrystalline samples reflect some average of $H_{\mathrm{c2}}^{\|}(T)$ (for field parallel to the $ab$ plane) and $H_{\mathrm{c2}}^{\bot}(T)$ (for field perpendicular to the $ab$ plane). The large slope could mainly reflect $H_{\mathrm{c2}}^{\|}(T)$ which is expected to be obviously steeper than $H_{\mathrm{c2}}^{\bot}(T)$ near $T_\mathrm{c}$ because of the insulating spacer layers in the 12442-type structure. Indeed, one sees a large gap between $H_{\mathrm{c2}}(T)$ and $H_{\mathrm{irr}}(T)$ curves, consistent with the enhanced anisotropy due to weak interlayer coupling. The exceptionally high value of the initial $H_{\mathrm{c2}}(T)$ slope, which suggests very short superconducting coherence lengths, deserves further investigations.

\begin{figure}
  \center
  \includegraphics[width=8cm]{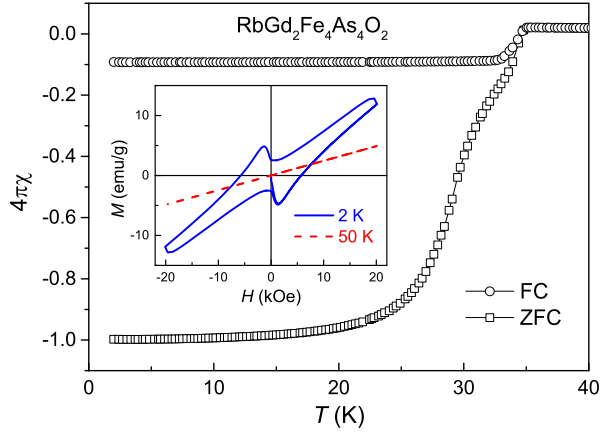}
  \caption{\label{magnetic} Superconductivity in RbGd$_2$Fe$_4$As$_4$O$_2$ evidenced by the dc magnetic susceptibility measured at $H$ = 10 Oe in field-cooling (FC) and zero-field-cooling (ZFC) modes. Note that the data were corrected by removing the demagnetization effect. The left inset shows the isothermal magnetization at 2 and 50 K.}
\end{figure}

Figure~\ref{magnetic} shows the temperature dependence of dc magnetic susceptibility [scaled by 4$\pi\chi(T)$] under a magnetic field of $H$ = 10 Oe. A strong diamagnetic transition occurs at $T_{\mathrm{c}}^{\mathrm{onset}}$ = 35.0 K for both field-cooling (FC) and zero-field-cooling (ZFC) data. The volume fraction of magnetic shielding is almost 100\% at 2 K after a demagnetization correction. Although the volume fraction of magnetic repulsion is greatly reduced because of magnetic-flux pinning effect, it is still appreciably high ($\sim$10\%), indicating that the title compound is responsible for the superconductivity. Shown in the inset of Fig.~\ref{magnetic} is the isothermal magnetization curves at 2 and 50 K. Above $T_{\mathrm{c}}$, it shows a paramagnetic behavior, which is mainly due to the Curie-Weiss paramagnetism of Gd$^{3+}$ spins ($S$ = 7/2). In the superconducting sate, a magnetic hysteresis loop is evidently superposed, from which a lower critical field of about 500 Oe is obtained. Since the corresponding $H_{\mathrm{c2}}$ value is much higher, RbGd$_2$Fe$_4$As$_4$O$_2$ is thus an extremely type-II superconductor.

\begin{figure}
	\center
	\includegraphics[width=8cm]{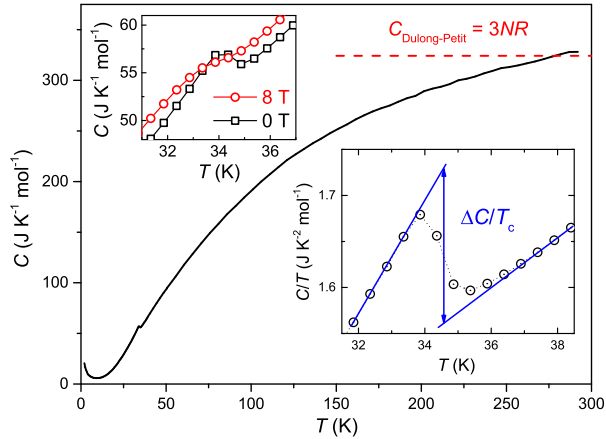}
	\caption{\label{CT} (a) Temperature dependence of specific heat for RbGd$_2$Fe$_4$As$_4$O$_2$. The horizontal dashed line represents the Dulong-Petit limit. The upper-left inset compares the specific-heat data around $T_{\mathrm{c}}$ at zero field and 8 T. The lower-right inset plots $C/T$ vs. $T$, from which the specific-heat jump and the thermodynamic transition temperature can be estimated.}
\end{figure}

Bulk superconductivity in RbGd$_2$Fe$_4$As$_4$O$_2$ is further confirmed by heat capacity measurement. As shown in Fig.~\ref{CT}, there is a specific-heat jump at $T_{\mathrm{c}}$ = 35 K. Under an 8-T magnetic field, the specific-heat anomaly tends to be smeared out and, $T_{\mathrm{c}}$ decreases slightly by $\sim$0.5 K, consistent with the above magnetoresistance measurement. The thermodynamic transition temperature is 34.55 K, based on an entropy-conserving construction (see the lower-right inset). The $\Delta C/T_{\mathrm{c}}$ value is 170$\pm$10 mJ K$^{-2}$ mol$^{-1}$, slightly higher than that (150 mJ K$^{-2}$ mol$^{-1}$) of KCa$_2$Fe$_4$As$_4$F$_2$\cite{wzc} but, a little lower than that (186 mJ K$^{-2}$ mol$^{-1}$) of CsEuFe$_4$As$_4$\cite{CsEu1144} with almost identical $T_{\mathrm{c}}$. Note that the Sommerfeld coefficient $\gamma$ in the normal state cannot be estimated reliably from the $C(T)$ data, primarily because of the high $T_\mathrm{c}$ value and the Gd magnetism incorporated. The dimensionless parameter $\Delta C$/($\gamma T_{\mathrm{c}}$) cannot be obtained accordingly. If assuming a weak-coupling scenario that satisfies the Bardeen-Cooper-Schrieffer (BCS) relation $\Delta C$/($\gamma T_{\mathrm{c}}$) = 1.43, conversely, one may estimate that the $\gamma$ value is about 120 mJ K$^{-2}$ mol$^{-1}$. If this is the case, the electronic specific heat achieves $\sim$36 J K$^{-1}$ mol$^{-1}$ at 300 K, which accounts for the large room-temperature specific heat beyond the Dulong-Petit limit. The low-temperature upturn in $C(T)$ comes from the Gd$^{3+}$ moments that could be ordered at lower temperatures. Besides, the upward shift of $C(T)$ under magnetic field, shown in the upper-left inset of Fig.~\ref{CT}, is due to field-induced reorientation of the Gd$^{3+}$ spins.

\begin{figure}
	\center
	\includegraphics[width=8cm]{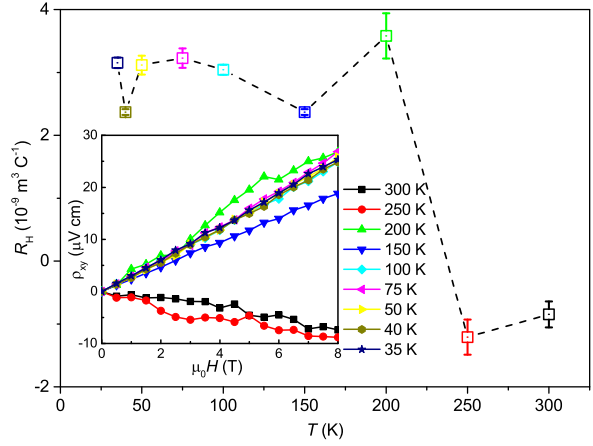}
	\caption{\label{Hall} Hall coefficient measurement for the RbGd$_2$Fe$_4$As$_4$O$_2$ polycrystalline sample. The inset shows field dependence of the Hall resistivity $\rho_{xy}$.}
\end{figure}

With formal valence state of Rb$^{1+}$Gd$_{2}^{3+}$Fe$_{4}^{2+}$As$_{4}^{3-}$O$_{2}^{2-}$, there remains one less positive charge which reflects a self hole doping in the present system. To verify this, we measured Hall coefficient ($R_\mathrm{H}$) for the RbGd$_2$Fe$_4$As$_4$O$_2$ sample. As is seen in Fig.~\ref{Hall}, the $R_\mathrm{H}$ values are positive below 200 K, indicating dominant hole conduction in the temperature region (estimation of the hole concentration is difficult because of multiple energy bands). In between 200 and 250 K, however, there is a sign change, in contrast with positive $R_\mathrm{H}$ below 300 K for KCa$_2$Fe$_4$As$_4$F$_2$\cite{wzc}. The sign change in $R_\mathrm{H}(T)$ resembles the case of the first hole-doped FeSC (La$_{1-x}$Sr$_{x}$)OFeAs\cite{whh}, which is probably due to a multi-band effect.

As is known, the $T_\mathrm{c}$ record of hole-doped FeSCs is 38 K\cite{johrendt} and to the best of our knowledge, in hole-doped \emph{oxyarsenide} FeSCs, the highest $T_\mathrm{c}$ of 25 K remains unchanged for years\cite{whh}, which was obtained in the 1111-type (La$_{1-x}$Sr$_{x}$)OFeAs containing single separate single Fe$_2$As$_2$ layer. Here, the $T_\mathrm{c}$ value of 35 K in RbGd$_2$Fe$_4$As$_4$O$_2$ represents a new record for hole-doped oxyarsenide FeSCs, which seems to be relevant to the double Fe$_2$As$_2$ layers. Thus it is of great interest to see whether the $T_\mathrm{c}$ value in 12442-type materials can achieve or even surpass the record of 38 K for the future.
\
\ack
This work was supported by the National Science Foundation of China (Nos. 11474252 and 11190023) and the National Key Research and Development Program of China (No. 2016YFA0300202)

\section*{References}

\bibliographystyle{12442}
\bibliography{12442}

\end{document}